# Operationalising the Definition of General Purpose AI Systems: Assessing Four Approaches

Risto Uuk[1], Carlos Ignacio Gutierrez[1], Alex Tamkin[2]
[1] Future of Life Institute
[2] Stanford University

## Abstract

The European Union's Artificial Intelligence (AI) Act is set to be a landmark legal instrument for regulating AI technology. While stakeholders have primarily focused on the governance of fixed purpose AI applications (also known as narrow AI), more attention is required to understand the nature of highly and broadly capable systems. As of the beginning of 2023, several definitions for General Purpose AI Systems (GPAIS) exist in relation to the AI Act, attempting to distinguish between systems with and without a fixed purpose. In this article, we operationalise these differences through the concept of "distinct tasks" and examine four approaches (quantity, performance, adaptability, and emergence) to determine whether an AI system should be classified as a GPAIS. We suggest that EU stakeholders use the four approaches as a starting point to discriminate between fixed-purpose and GPAIS.

## 1. Introduction

The European Union (EU) has made significant efforts toward artificial intelligence (AI) governance to address this technology's potential risks and harms. This was demonstrated with the release of the draft AI Act, a landmark legal instrument, in April of 2021. Past success with the General Data Protection Regulation has proved the EU's ability to impact emerging technology governance around the world, making it a leader for other governments and entities pursuing similar goals in AI governance (De Ville & Gunst, 2021; Siegmann & Anderljung, 2022).

One of the foundational underpinnings of the draft version of the EU AI Act is the notion of a system's purpose. Many AI systems have an identifiable fixed purpose that can be used to classify them as low or high risk to health, safety, or fundamental rights. However, general purpose AI systems (GPAIS), as the EU calls them, might defy this presumption. These systems lack a particular intended purpose and can be adapted by users or autonomously serve an indeterminate number of objectives with varying levels of risk.

The goal of this article is to evaluate four approaches for identifying GPAIS. In particular, it focuses on thresholds in the literature that can help distinguish what qualifies as a GPAIS. This piece is divided into three sections. The first section summarises existing approaches to define GPAIS by the EU and external actors. It finds that a crucial common denominator between proposals is the need to clarify and operationalise what can be considered a unique purpose, herein discussed as "distinct tasks." The second section discusses four approaches (quantity, performance, adaptability, and emergence) that stakeholders can use to make this differentiation. The last section examines the overarching role that these approaches should play in the EU's governance of AI.

## 2. Defining GPAIS in an EU Context

To scope the AI Act adequately, the EU has undergone a process of defining foundational terms. The definition of the term "AI" itself has been widely debated. On the one hand, certain interest groups have advised against a definition that excludes relevant systems (Bryson, 2022). On the other hand, member states advocated for a narrower definition that was adopted by the Council in December 2022 (Council of the European Union, 2022). Concurrently, the European Parliament is considering both narrower and broader visions for defining AI (European Parliament, 2022).

The term GPAIS has not received as much attention as the definition of AI. Although this concept was rarely employed prior to the EU's development of the AI Act, it is now a central idea for describing systems without a fixed purpose (Gutierrez et al., 2022a). As a technology, GPAIS are important because they include an increasingly powerful set of systems that are being deployed widely, such as ChatGPT, Bard, and Bing Chat. The EU has proposed a definition for GPAIS that has faced significant criticism from external parties of being too broad in scope. Many actors engaged in this process have proposed alternative definitions to better identify systems unconstrained by a fixed purpose (Morrison, 2022). The following table contains a summary of proposed GPAIS definitions in the context of the AI Act.

Table 1: GPAIS definitions relevant to the EU context

| Draft EU Position (Council of the European Union, 2022) | AI system that - irrespective of how it is placed on the market or put into service, including as open source software - is intended by the provider to perform generally applicable functions such as image and speech recognition, audio and video generation, pattern detection, question answering, translation and others; a general purpose AI system may be used in a plurality of contexts and be integrated in a plurality of other AI systems. |
|---|---|
| Gutierrez et al., 2022 | An AI system that can accomplish or be adapted to accomplish a range of distinct tasks, including some for which it was not intentionally and specifically trained. |
| Gahntz & Pershan, 2022 | AI systems that are provided without a specific intended purpose; instead, they can serve a large number of purposes, including |

| | purposes not foreseen or declared by their original providers. |
|---|---|
| Engler & Renda, 2022 | AI systems characterised by their training on especially large datasets to perform many tasks, making them particularly well-suited for adaptation to more specific tasks through transfer learning. |
| Campos & Laurent, 2023 | AI systems that can accomplish a range of distinct valuable tasks, including some for which it was not specifically trained. |
| Moës, 2022 | • Preferred definition: General purpose AI systems are AI systems that score above x% on the EU standardised testing suite for generality administered by the European Benchmarking Institute.<br>• OK/temporary definition: General purpose AI systems are AI systems that can be reasonably foreseen to carry out a broad range of tasks (e.g., ≥ 10) from the EU official list of tasks without substantial modification. |

The definitions suggested by external actors contrast with the one in the draft AI Act. The wording in the existing EU proposal is potentially over-inclusive of a wide range of technologies under GPAIS, possibly including simple methods like linear regression. Meanwhile, the five proposals offer a concrete perspective on factors that distinguish GPAIS. A key element in all of them is the notion that the technology performs different functions (according to the EU), tasks (according to Gutierrez et al., Campos & Laurent, Engler & Renda, and Möes), or purposes (according to Gahntz & Pershan). Henceforth, this article will refer to these terms as "distinct tasks." That is to say, a GPAIS is distinguished from narrow AI systems based on its versatility at completing distinct tasks, and a lack of a fixed purpose. Clearly distinguishing what constitutes a distinct task is necessary in order to identify the key characteristics of this technology and strengthen a system's case for "generality."

When the AI Act becomes regulation, whatever form the final definition of GPAIS takes, stakeholders will require clarification as to what constitutes a distinct task. This will give the term precision and separate these systems from their fixed purpose counterpart. This is ever more important considering the obstacles that stakeholders face in applying this definition. Such was the finding of a survey in which 13% of EU start-ups were not sure if their products could be considered GPAIS based on the EU's draft definition (Andreas Liebl & Till Klein, 2022). Moreover, 45% affirmatively stated that their systems are GPAIS, even though the draft definition lacks clear inclusion or exclusion criteria.

As will be discussed in the following section, there are several ways to differentiate between what a GPAIS can do. This article compiles them into the question of: what approaches could be used to discriminate between "distinct tasks?"

# 3. Approaches for "Distinct Tasks"

The concept of "distinct tasks" is a critical element with which to characterise the versatility of AI systems. As European regulators finalise the AI Act, they will face the challenge of clarifying how stakeholders classify the status of their products in terms of GPAIS. In this section, we offer four approaches for determining if an AI system should be considered a GPAIS.

One approach to defining a GPAIS is based on the quantity of distinct tasks it can in principle perform. Another is to judge it by its effectiveness in practice at accomplishing those distinct tasks. A third approach is measuring an AI system's adaptability, or how well it can learn to achieve new distinct tasks. Finally, a fourth approach is the degree to which AI systems emergently develop the ability to execute distinct tasks.

Each approach has strengths and weaknesses that should inform concrete recommendations for how to define GPAIS in a manner that is feasible and effective. In all cases, we urge stakeholders to recognise how these approaches cope with the need for EU stakeholders to apply criteria that are at least:

1. **Practical:** Distinguishing a GPAIS is straightforward, actively minimises uncertainty, and is not overly burdensome.
2. **Flexible:** Any criteria is applicable to a wide range of existing and expected capabilities. This incorporates fields such as natural language processing, computer vision, speech, and robotics, among others.
3. **Future-proof:** Our conception of GPAIS will change as the technology develops. Any methodology to define it must be susceptible to proactive action that addresses upcoming advances or sufficiently reactive to cover unexpected technological leaps. For policymakers, both would be preferable.

## 3.1. Quantity

The first approach to identifying a GPAIS is based on the number of distinct tasks it could be applied to. Moës (2022) suggests a two-pronged proposal to identify GPAIS using this line of thought. Firstly, the EU would generate a list enumerating the distinct tasks that an AI system can perform. Secondly, a threshold number of tasks is established to separate fixed purpose systems from GPAIS. For this proposal to work, the list of distinct tasks need to be credible. Because an official EU list does not exist, the EU standardisation bodies could play a role in its creation. Meanwhile, there are a few alternatives to consider, such as:

Table 2: Alternative lists for identifying the number of distinct tasks

| O*NET task statements (*Task Statements - O*NET 20.1 Data Dictionary at O*NET Resource Center*, n.d.) | Map of 19,500 tasks associated with occupations. For example, a task related to electrical engineering can include the operation of computer-assisted engineering, software design software, or the use of equipment to perform engineering tasks. |
|---|---|

| O*NET intermediate work activities (*IWA Reference - O*NET 20.1 Data Dictionary at O*NET Resource Center*, n.d.) | List of 300 intermediate work activities divided into broad categories such as analysing data, assisting others, coaching, communicating, and many more. |
|---|---|
| The Beyond the Imitation Game benchmark (BIG-bench) (*BIG-Bench*, 2023) | Contains 213 tasks that are part of the Beyond the Imitation Game benchmark (BIG-bench). Each subdirectory contains a single benchmark task. It is intended for the evaluation of capabilities from large language models. |
| The AI Index Report 2023 framework (Maslej et al., 2023) | Framework to describe AI tasks according to various broad categories (and their subcategories) such as computer vision (image and video), language, speech, reinforcement learning, hardware, environment, and AI for science. |

Stakeholders considering this approach are faced with the practical issue of designing appropriate and justified thresholds. In all cases, they should minimise the burden of distinguishing a GPAIS. An issue to keep in mind is that any selected method within this scope must also take into consideration the performance an AI system achieves on various distinct tasks. Equally relevant is that stakeholders must be cognisant of the incentives placed on developers when generating a numerical threshold, as this could be used to avoid restrictions or enhanced scrutiny.

There are several resources that may help EU stakeholders assemble such a list of distinct tasks covering many different capabilities. A number of these lists are capability- and modality-limited such as BIG-Bench, which is dedicated to language models. For example, BIG-Bench includes specific and distinct tasks such as: "Answer questions designed to probe social biases," "Evaluate the result of a random Boolean expression," and "Identify legal moves in the given chess position." Other lists, such as the AI Index, cover a range of capabilities and modalities besides language, namely computer vision and reinforcement learning. Similarly, the O*NET task statements are expansive, as they include a wide range of tasks covering capabilities required for existing occupations. This list is complemented by the O*NET intermediate work activities, which contains increasingly general groupings.

However, there are several drawbacks to the Quantity approach. For example, one major challenge is to future-proof the approach by keeping these lists updated. This is important because new tasks are constantly being created with the development of new products and technologies. EU stakeholders will find that this requires specialised human resources regularly dedicating bandwidth to identify and map new distinct tasks. This could be done by EU benchmarking authorities and experts in the Artificial Intelligence Board. Alternatively, a framework generalising the difference between tasks could serve a practical purpose by identifying tasks that are not known ahead of time.

In addition, there are other procedural questions to consider when assembling such lists of tasks. For one, any list should minimise the degree of overlap between distinct tasks so that tasks to avoid duplication. Secondly, distinct tasks ought to be adequately abstract to be meaningful, but not so abstract that they are difficult to understand or apply. In the case of O*NET, for example, many tasks could be grouped into a single larger category, including both "recommending" economic policies and "advising" on economic policies. In contrast, on the O*NET list of task statements these tasks would be considered distinct.

## 3.2. Performance

A second approach to discriminating systems relies on measuring how effective they are in completing distinct tasks. It might not be sufficient to ask how many tasks a model could theoretically perform, because that casts too broad a net: for example, a rudimentary autocomplete system could in theory be used to write the rest of a complex report, but it would do so poorly. Instead, this approach focuses on discriminating between systems based on how well they perform different tasks, as the most relevant systems for regulation will not merely be applicable in theory to many tasks but actually perform those tasks meaningfully well.

A key method of performance measurement in machine learning is using benchmarks, which contrast a system's ability to complete a task relative to others. However, there are practical issues to relying on benchmarks as a means to discriminate if a distinct task is achieved. For one, EU stakeholders must decide which metrics are most appropriate. They can rely on the vast number of benchmarks available in the field, but it is important to consider that many remain unused or, in some cases, few advances are observed (Ott et al., 2022). Furthermore, stakeholders should examine whether a benchmark is designed in a manner that aligns with the EU's goals, since benchmark designs can contain misaligned objectives or measure targets in misleading ways (Raji et al., 2021).

One important example from the field is the holistic evaluation of language models (HELM) approach. It states that, among other things, performance measurement needs to be multi-metric because societally beneficial systems reflect many different values (Liang et al., 2022). Overall, setting a threshold for what constitutes a "satisfactory" task performance is a critical, achievable challenge. However, when doing so policymakers need to consider how developers may seek to avoid scrutiny of their products by misrepresenting their system's capabilities.

As a comparison tool, benchmarks exist for a wide array of distinct tasks spanning many capabilities. For language models, they can measure question answering, missing word completion, and a growing scholarship dedicated to quantifying these models' negative externalities (Minaee et al., 2021; Paperno et al., 2016; Rauh et al., 2022). Benchmarks for tasks in other modalities (e.g., audio, video, speech, among others) are publicly available (*Papers with Code - Machine Learning Datasets*, n.d.). In this sense, EU stakeholders can take advantage of this diverse supply of benchmarks as a means of ensuring their flexibility in characterising systems that classify as GPAIS.

For the foreseeable future, developers will compare their systems with competitors through benchmarks. In parallel, benchmarks will be created based on the availability and performance of AI systems. The EU can take advantage of this market in order to distinguish GPAIS from their fixed purpose counterparts by adapting existing benchmarks for its purposes, as well as developing their own. With this benchmark-based approach with performance as a

GPAIS determinant, policymakers still face the challenge of reaching a consensus for a large range of distinct tasks.

### 3.3. Adaptability

The third approach to characterise GPAIS is assessing their ease in being adapted to perform new distinct tasks. While both GPAIS and fixed purpose systems can execute at least one task, as the number of tasks grows, we can distinguish technologies by their ability to accomplish additional tasks they are applied to. For fixed purpose systems, the performance of a list of tasks is relatively step-wise: the system is either able to generate an output for a task to a degree, or not at all. For instance, a facial recognition system will generate an output when it is provided with visual information of a person's face. This system is not useful unless given an image, and its ability to recognise other objects is limited by its training. By contrast, GPAIS could learn to perform new tasks such as classifying new kinds of objects. In practice, this adaptation is often done by conditioning and priming the GPAIS with examples of a description of the task, or by modifying or fine-tuning its parameters (Bommasani et al., 2022).

As more AI systems are developed with general-purpose capabilities, one way EU stakeholders can identify them in practice with this approach is by their few-shot or zero-shot learning ability – meaning that they perform well on some tasks even when exposed to few or no examples or instructions. Large language models are already considered good few-shot learners and researchers have even conducted additional improvements in these models by adding "Let's think step by step" in the prompt (Kojima et al., 2022).

Adaptability is a characteristic present throughout the landscape of AI system capabilities, making it a flexible or technology-neutral approach. In addition to language systems, zero- to few-shot learning is an ability present in systems dedicated to images, videos, and interactions with virtual and physical environments (Reed et al., 2022; Stooke et al., 2021). Furthermore, adaptability is relevant because it often makes broader capabilities available even when GPAIS have ostensibly been scoped to a single task. An example of this is the AI Dungeon video game, which customised the language model GPT-3 for fantasy role-playing. By employing specific prompts, users were able to adapt a system originally limited to one purpose to another with all of GPT-3's capabilities, revealing strong adaptability to many distinct tasks through the use of an open-ended interface (Ganguli et al., 2022). Narrow AI systems could not be adapted to a new distinct task in this manner.

Future systems may become increasingly good zero-shot reasoners (and already are to some extent). This might mean that, over time, a GPAIS will not only easily adapt to new distinct tasks, but may also perform them without much or any adaptation required. In effect, this is similar to how humans engage with new information and learn to adapt to novel environments. EU stakeholders contemplating adaptability as an inclusion criterion for GPAIS must be aware that not all systems without a fixed purpose are capable of completing distinct tasks in this manner. In other words, zero-shot learning can be accomplished for a subset of task types, and systems capable of this should be considered GPAIS.

### 3.4. Emergence

The final approach for determining a GPAIS is to discern a system's potential for developing emergent abilities that enable it to perform distinct tasks. Emergence is when "quantitative changes in a system result in qualitative changes in behavior" (Wei et al., 2022). This means that certain systems can develop distinct emergent task abilities as their amount of computation, parameters, or training dataset size grows (Wei et al., 2022). For example, a 1B parameter model may not be able to accomplish tasks that the same architecture of model with a 100B parameter count can. The same may be true for one trained with $10^{20}$ versus $10^{24}$ FLOPs (Wei et al., 2022).

Emergence is relevant for identifying GPAIS because it makes it impossible for a system to be scoped to a fixed task or set of tasks. Moreover, even after a model is trained, its creators and users may not be aware of all of its capabilities, and certain areas of competency may only be discovered when a specific type of input is provided (Ganguli et al., 2022). As of early 2023, there is evidence that at least large language models show emergent abilities. It is possible that with some architectural or algorithmic breakthroughs, some AI systems could experience emergence without needing to achieve the scale of current frontier systems.

Using emergence as a distinguishing property for GPAIS poses practical issues because qualifying systems cannot always be identified a priori. Although we observe that GPAIS can likely perform new distinct tasks when scaled-up, where more narrow systems cannot, it is necessary to gain a better understanding as to what factors distinguish those systems exhibiting emergent behaviour (Wei, 2022). Accordingly, while identification or reasonable prediction of emergence may lend support to classifying a system as a GPAIS, the absence of observed emergence alone may not be sufficient to disqualify a system from such a classification. Although we expect researchers to improve their grasp on what characterises emergence, it is currently too early to tell whether the use of this characteristic to distinguish GPAIS will stand the test of time. Despite this, insofar as the possibility of emergent distinct tasks might be foreseeable, this assessment should be done.

Based on available research, emergence is not a flexible characteristic for determining a system's GPAIS status, particularly in its absence. The evidence shows that so far emergence is largely manifesting in large language models. "There are more than 100 examples of emergent abilities that have already been empirically discovered by scaling language models such as GPT-3, Chinchilla, and PaLM" (Wei, 2022). These include knowledge of Hindu, modified arithmetic, causal judgment, geometric shapes, and physics questions. Other examples of emergent-like behaviour can be found in reinforcement-learning research outputs (Bauer et al., 2023) as well as in vision models (Caron et al., 2021).

Emergence is a salient attribute because it acknowledges that there may be new tasks lurking beneath the surface of current models that have yet to be discovered. Further work is required to determine whether this phenomenon is applicable to other types of AI systems. When observed, emergence is generally a good indication of a general purpose system, but because the opposite is not always true, it is not an ideal determinant.

## 4. Conclusion

A single consensus method for identifying GPAIS does not yet exist, and the current state of AI governance in the EU lacks a clear path for accomplishing this objective. Such clarity is critical considering the significant potential for harms of GPAIS and the need to include it within the scope of the AI Act's risk approach. The goal of this paper is to analyse potential approaches to better distinguish GPAIS from fixed purpose AI systems.

This paper contains four approaches that EU stakeholders should consider when operationalising the inclusion and exclusion criteria for a GPAIS. Each has advantages and weaknesses that need to be addressed based on the EU's ability to maximise the practicality, flexibility, and future-proofness of any selected approach. We propose that the EU consider each of the four approaches (quantity, performance, adaptability, and emergence). If an AI system surpasses a previously established and consented EU threshold for any of them, then a technology should be considered a GPAIS under the AI Act. However, while each approach alone could give reason to consider an AI system a GPAIS, but they are best used holistically. It is important not to include narrow systems in the criteria so that the risk management procedures prescribed for these systems can be sufficiently robust, without being onerous for creators of simpler more predictable systems.

Moreover, it is necessary to recognise that the approaches in this document are a sample of the existing means to distinguish GPAIS. Our intention is to bring to light important ideas shared in the literature. Similarly, we should consider the implications of a given method for identifying GPAIS. For instance, policymakers must assess the potential incentives an approach can create and prevent negative outcomes. Otherwise, stakeholders can "game" regulation by intentionally failing to meet thresholds for the number of tasks or their performance effectiveness on standardised tests. If this were to happen, it would enable technologies to escape necessary scrutiny and increase the risk to individuals, groups, nations, and the planet. While it is challenging to operationalise a definition for GPAIS, it is a critical government task and we hope the considerations in this paper can play a role in minimising the risks of this technology.

## Acknowledgements

We would like to thank the following individuals who provided valuable feedback to this research article: Anthony Aguirre, Connor Dunlop, Irene Solaiman, Landon Klein, Jack Clark, Jared Mueller, Marius Hobbhahn, Mark Brakel, Neil Natarajan, Richard Mallah, Siméon Campos, and Yonadav Shavit.